\documentclass[twocolumn,showpacs,preprintnumbers,amsmath,amssymb,superscriptaddress]{revtex4}

\usepackage{graphicx}
\usepackage{dcolumn}
\usepackage{bm}
\usepackage{color}

\begin{document}


\title{Symmetry-lowering lattice distortion at the spin-reorientation in MnBi single crystals}

\author{Michael A. McGuire}
\email{mcguirema@ornl.gov}
\affiliation{Materials Science and Technology Division, Oak Ridge National Laboratory, Oak Ridge, Tennessee 37831, USA}
\author{Huibo Cao}
\affiliation{Quantum Condensed Matter Division, Oak Ridge National Laboratory, Oak Ridge, Tennessee 37831, USA}
\author{Bryan C. Chakoumakos}
\affiliation{Quantum Condensed Matter Division, Oak Ridge National Laboratory, Oak Ridge, Tennessee 37831, USA}
\author{Brian C. Sales}
\affiliation{Materials Science and Technology Division, Oak Ridge National Laboratory, Oak Ridge, Tennessee 37831, USA}

\date{\today}

\begin{abstract}
Structural and physical properties determined by measurements on large single crystals of the anisotropic ferromagnet MnBi are reported. The findings support the importance of magneto-elastic effects in this material. X-ray diffraction reveals a structural phase transition at the spin reorientation temperature $T_{SR}$ = 90 K. The distortion is driven by magneto-elastic coupling, and upon cooling transforms the structure from hexagonal to orthorhombic. Heat capacity measurements show a thermal anomaly at the crystallographic transition, which is suppressed rapidly by applied magnetic fields. Effects on the transport and anisotropic magnetic properties of the single crystals are also presented. Increasing anisotropy of the atomic displacement parameters for Bi with increasing temperature above $T_{SR}$ is revealed by neutron diffraction measurements. It is likely that this is directly related to the anisotropic thermal expansion in MnBi, which plays a key role in the spin reorientation and magnetocrystalline anisotropy. The identification of the true ground state crystal structure reported here may be important for future experimental and theoretical studies of this permanent magnet material, which have to date been performed and interpreted using only the high temperature structure.
\end{abstract}

\maketitle

\section{Introduction}

Due to its high Curie temperature and strong magnetic anisotropy, MnBi has attracted increasing attention in the pursuit of rare-earth-free permanent magnets (see for example the review by Poudyal and Liu \cite{Poudyal-2013}). Advances in the understanding of the properties and processing of this material continue, despite its relatively simple, NiAs-type crystal structure, and the fact that ferromagnetism in MnBi was first reported over a century ago \cite{Heusler-1904}. Key early studies were preformed by Guillaud \cite{Guillaud-1951a, Guillaud-1951b}. In the 1950's, this material was considered as a candidate to replace permanent magnets containing cobalt and nickel, and an energy product of 4.3 MGOe was achieved in the resulting ``Bismanol'' magnets \cite{Adams-1952, Adams-1953}. Subsequently, energy products as high as 7.7 MGOe have been reported \cite{Yang-2002}. MnBi has several promising aspects as a candidate replacement for rare earth magnets. These include: (1) relatively inexpensive components \cite{Coey-2011}, (2) high ordered moment and saturation magnetization of 0.58 MA m$^{-1}$ or about 3.5 $\mu_B$ per Mn at room temperature \cite{Roberts-1956}, (3) ferromagnetism that persists to 630 K (about 40 K higher than Nd$_{2}$Fe$_{14}$B) \cite{Guillaud-1951a}, (4) large and uniaxial magnetocrystalline anisotropy energy near 1 MJ m$^{-3}$ at room temperature (moments along the c-axis), which increases upon heating above room temperature \cite{Williams-1956, Albert-1961, Chen-1974, Guo-1992}. The observed increase in magnetic anisotropy and coercivity with increasing temperature is perhaps the most interesting, unique, and potentially important property of MnBi.

The ferromagnetism in MnBi vanishes abruptly upon heating above 630 K \cite{Guillaud-1951a, Guillaud-1951b}.  Importantly, the phase transition that occurs near 630 K is not a typical magnetic transition. This is not technically the Curie temperature, and the transition is not directly driven by magnetism. This transitions has been identified as a peritectic decomposition of MnBi. Heikes noted that the decomposition products contain a ferromagnetic phase with a Curie temperature well below the decomposition temperature of MnBi \cite{Heikes-1955}. It has since been determined that near 630 K MnBi decomposes into Mn$_{1.08}$Bi and Bi \cite{Chen-1973}, both of which are paramagnetic at this temperature. The Mn rich phase is an orthorhombic variant of the NiAs-structure with the excess Mn occupying trigonal prismatic interstices similar to those occupied by the Bi atoms \cite{Cenzual-1991}. Since MnBi is ferromagnetic and the decomposition products are paramagnetic (at the peritectic temperature), application of strong magnetic field has been shown to stabilize MnBi to higher temperatures, up to about 650 K in a 10 Tesla magnetic field \cite{Liu-2005, Onogi-2007}.

Synthesis of high quality single phase samples and in particular large single crystals of MnBi is complicated by the peritectic nature of this reaction, the relatively low temperature at which it occurs (630 K), and the nearness to the eutectic temperature (535 K). MnBi can be grown from a melt with excess Bi, as first described by Adams et al. \cite{Adams-1952} By cooling the melt in a magnetic field, textured composites with coaligned MnBi crystallites embedded in a Bi matrix have been obtained and studied \cite{Morikawa-1998, Liu-2005}. For permanent magnet applications, fine-grained, polycrystalline samples are usually desired. These have been produced from melt-spun and/or mechanically milled material \cite{Guo-1990, Saha-2002}, and by magnetic separation of MnBi from excess Bi in samples produced by powder-metallurgical routes \cite{Yang-2002}.

Magnetic measurements show that the magnetocrystalline anisotropy of MnBi passes through zero and the coercivity of MnBi powders vanishes near 90 K \cite{Albert-1961, Chen-1974, Yoshida-2001, Yang-2001}. These observations are associated with the reorientation of the Mn magnetic moment from parallel to perpendicular to the crystallographic c-axis upon cooling. The spin reorientation temperature, $T_{SR}$, denotes the temperature at which magnetic moment develops a c-component upon heating and the coercivity begins to increase. The rotation of the moments has been oberved in neutron diffraction from MnBi powders \cite{Roberts-1956, Andersen-1967, Yang-2002} and also nuclear magnetic resonance (NMR) experiments \cite{Hihara-1970}. The NMR results from powders \cite{Hihara-1970} and magnetization measurements on single crystals \cite{Chen-1974} show the rotation away from the \textit{c}-axis upon cooling begins near 140 K, increases gradually upon further cooing to $T_{SR}$ = 90 K, and then abruptly completes, with the moment flopping discontinuously into the \textit{ab}-plane. However, neutron diffraction from MnBi powder suggests the moments are not fully in-plane below 90 K \cite{Roberts-1956, Yang-2002}.

In the presence of significant magneto-elastic coupling, a response in the crystal lattice to the reorientation of the magnetic moments is expected. Signatures of this coupling have been reported in the temperature dependence of the lattice constants of MnBi \cite{Yoshida-2001, Koyama-2008, Yang-2011}. While the hexagonal NiAs structure type describes the powder x-ray diffraction results at all temperatures, anomalies in both the a and c lattice parameters occur at temperatures near the spin reorientation. Recent first-principles calculations have reproduced these magneto-elastic effects, and linked the spin-reorientation to anisotropic thermal expansion \cite{Zarkevich-2014}, and in particular to resulting changes in the anisotropic pairwise exchange interactions between Bi p-states \cite{Antropov-2014}.

Most of the experimental studies performed on MnBi to date have used polycrystalline material. Single crystals are better suited for the detailed study of intrinsic and anisotropic properties. Here we report the results of our detailed investigation of the structural and physical properties of single crystal MnBi, including x-ray and neutron diffraction, magnetization, heat capacity, and electrical resistivity measurements. In addition to elucidating the intrinsic properties of MnBi, we find significant differences between the behavior of the single crystals and the reported behavior of polycrystalline MnBi in the literature, in particular regarding the spin reorientation. Our x-ray diffraction results reveal a symmetry-lowering structural distortion occurs at $T_{SR}$, and suggest that MnBi adopts an orthorhombic structure at low temperature, similar to Mn$_{1.08}$Bi but without the excess interstitial Mn. This transition is extremely sensitive to strain, and is not observed in powders gently ground from the crystals. The structure change is accompanied by a thermal anomaly which responds strongly to applied magnetic field, and a sharp decrease in electrical resistivity.  Examination of the anisotropic atomic displacement parameters determined from neutron diffraction analysis shows that the Bi vibrations along the $c$-axis are enhanced with increasing temperature above the spin reorientation, suggesting a relationship with the well-documented increase in magnetocrystalline anisotropy energy with temperature.

\section{Experimental Details}

\begin{figure}[h!]
\begin{center}
\includegraphics[width=3.0in]{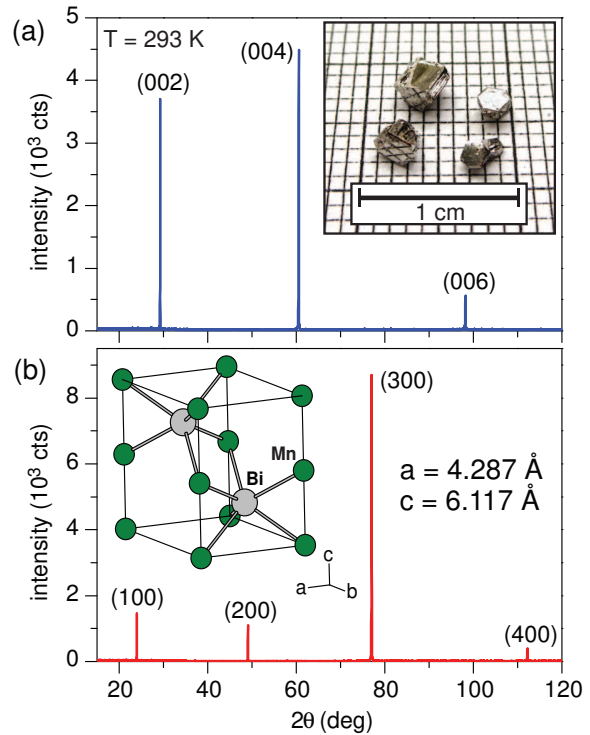}
\caption{\label{fig:crystals}
X-ray diffraction patterns from MnBi single crystal faces at room temperature. (a) Diffraction from a face perpendicular to the \textit{c}-axis. (b) Diffraction from a face perpendicular to the \textit{a}-axis. The inset in (a) shows a photograph of typical MnBi crystals. The inset in (b) shows the room temperature hexagonal crystal structure of MnBi. The lattice constants determined from the reflections shown in the figure are listed in (b).
}
\end{center}
\end{figure}

Starting materials for the crystal growths were Bi shot (Cominco American, 99.9999\%) and Mn pieces (Alfa Aesar, 99.95\%), and the crystal were grown from a flux with excess Bi based on the published binary phase diagram \cite{Mn-Bi-PD}. The Mn pieces were lightly ground into a powder and 0.65 g of this material was then immediately combined with 35 g of Bi and loaded into an alumina crucible. The 10 mL crucible, covered with an inverted 10 mL ``catch'' crucible half-filled with quartz wool was sealed under vacuum inside a silica ampoule. The ampoule was heated to 1000 $^{\circ}$C at 1 $^{\circ}$C/min, held for 24 h, cooled to 440 $^{\circ}$C at 1 $^{\circ}$C/min, held at 440 $^{\circ}$C for 1 h, and then cooled to 275 $^{\circ}$C at 0.4 $^{\circ}$C/h. At 275 $^{\circ}$C, the excess Bi flux was centrifuged into the catch crucible. Several mm-cm sized MnBi single crystals were produced via this method, some of which are shown in Fig. \ref{fig:crystals}a. The largest crystal obtained weighed close to 2 g. When exposed to air for hours, crystals developed a visible patina. X-ray diffraction from crystals which had been powdered and then exposed to air for more than 12 hours indicated slow decomposition into MnBi, MnO$_2$, and Bi.

X-ray diffraction measurements were conducted with a PANalytical X'Pert Pro MPD diffractometer, equipped with an incident beam monochromator (Cu K$\alpha_1$ radiation) and an Oxford PheniX closed-cycle helium cryostat. Single crystal neutron diffraction was performed at the HB-3A four-circle diffractometer at the High Flux Isotope Reactor at Oak Ridge National Laboratory. A neutron wavelength of 1.003 {\AA} was used from a bent perfect Si-331 monochromator \cite{HB3A}. The data were collected at the selected temperatures of 5 K, 50 K, 70 K, 90 K, 110 K,  130 K, 150 K, 200 K, 300 K, 350 K, 400 K, and 450 K.  At each temperature, more than 330 reflections were collected and used for the refinements.   Refinements of the X-ray and neutron diffraction data were performed using FullProf \cite{FullProf}. dc magnetization measurements were performed using a Quantum Design Magnetic Properties Measurement System and ac measurements were made using a Quantum Design Physical Properties Measurement System, which was also used for the resistivity and heat capacity measurements. Spot-welded Pt leads were used for electrical contacts.

\section{Results and Discussion}

\subsection{X-ray diffraction}

\begin{figure}
\begin{center}
\includegraphics[width=3.0in]{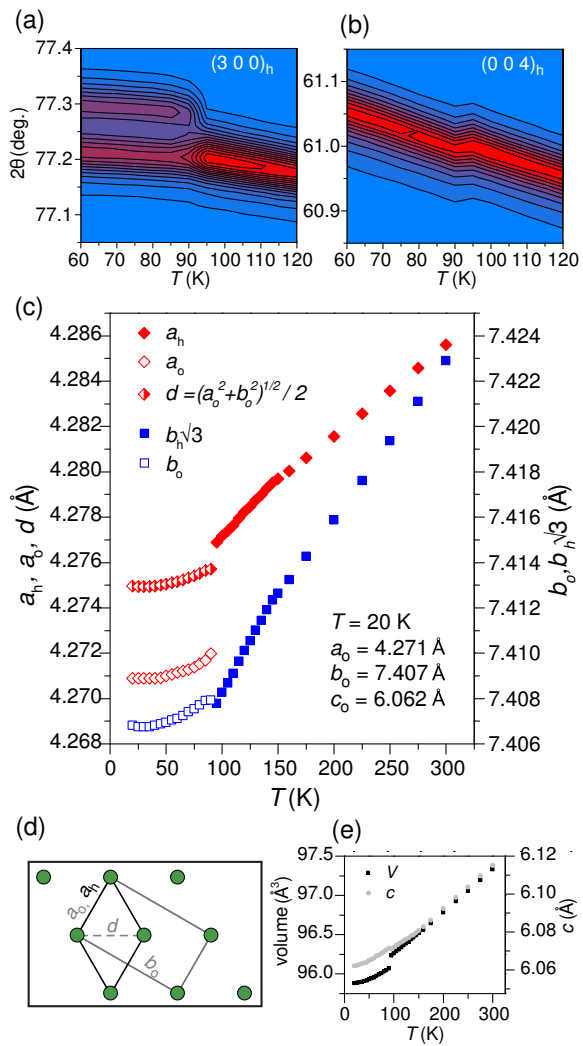}
\caption{\label{fig:xray}
Results of temperature dependent x-ray diffraction measurements from indexed faces of MnBi single crystals. (a,b) Contour plots of intensity vs diffraction angle and temperature for the 300 and 004 reflections (h denotes the hexagonal unit cell). (c) In-plane lattice constants vs temperature for the high temperature hexagonal and low temperature orthorhombic (denoted by subscript o) structures. (d) The relationship between the high temperature hexagonal and low temperature orthorhombic unit cells. (e) c-axis length and primitive unit cell volume vs temperature (the volume of the C-centered orthorhombic cell is twice the value plotted).
}
\end{center}
\end{figure}

A photograph of typical crystals used in this study is shown in Fig. \ref{fig:crystals}a. Diffraction from powdered crystals at room temperature confirmed them to be NiAs-type MnBi (space group $P$6$_3/mmc$). The crystal structure is shown in Fig. \ref{fig:crystals}b, along with the lattice constants determined from this room temperature data. Diffraction patterns, like those shown in Fig. \ref{fig:crystals}, from crystal facets were used to identify the hexagonal c-axis (typically a hexagonal face) and the a-axis (typically a rectangular face).

To examine the lattice response to the spin reorientation, diffraction patterns from crystal faces were collected at temperatures between 300 and 20 K. At 90 K, the sixfold symmetry in the \textit{ab}-plane is broken, as demonstrated by the splitting of the hexagonal 300 reflection in Fig. \ref{fig:xray}a. Such a distortion lowers the unit cell symmetry from primitive hexagonal to \textit{C}-centered orthorhombic. The temperature dependence of the lattice parameters determined from these measurements is shown in Fig. \ref{fig:xray}c and e. The relationship between the two unit cells is shown in Fig. \ref{fig:xray}d. For the orthorhombic structure, half of the face diagonal, labeled \textit{d} in the figure, is plotted as well in Fig. \ref{fig:xray}c. The sixfold symmetry in the hexagonal structure constrains these two distances to be the same, and so the divergence of \textit{a} and \textit{d} illustrates clearly the distortion. In addition, a change in the temperature dependence of the lattice parameters is observed in the hexagonal state near 140 K, where the spin reorientation onsets (see below).

There is a small expansion of the \emph{c}-axis upon cooling through 90 K (Fig. \ref{fig:xray}e). This is also apparent in the contour plot of the 004 reflection in Fig. \ref{fig:xray}b, which shows an increase in d-spacing (decrease in 2$\theta$) upon cooling at 90 K, the temperature below which the magnetic moment is confined to the \textit{ab}-plane. This suggests that magneto-elastic coupling produces a contraction of the lattice along the direction that the moment is directed. This is consistent with the behavior seen in the basal plane dimensions as well. Upon cooling through the spin reorientation, the basal plane contraction exceeds the \textit{c}-axis expansion, resulting in a sharp decrease in unit cell volume (Fig. \ref{fig:xray}e). The orthorhombic lattice parameters determined at 20 K from this data are listed in Fig. \ref{fig:xray}c.

The importance of the coupling between the properties of the crystal lattice and the magnetism in MnBi has been recently noted by Zarkevich et al. \cite{Zarkevich-2014}. They report a correlation between lattice parameters and magnetocrystalline anisotropy energy identified by density functional theory calculations. In particular, the calculated anisotropy is strongly dependent on \textit{a}. When the value of \textit{a} used in the calculations is increased, the preferred direction of the magnetic moment switches from in the \textit{ab}-plane to along the \textit{c}-axis. This is generally consistent with the lattice parameters reported here, where the \textit{ab}-plane of the orthorhombic phase (moments in the plane) is contracted relative to the hexagonal phase (moments out of the plane). First principles calculations reported by Antropov \textit{et al.} \cite{Antropov-2014} find that this is due to spin-orbit coupling and an unusual evolution of Bi$-$Bi interactions as the lattice constants change. Note that these calculations, and all others for MnBi reported to date, used the NiAs-structure at all temperatures.

Strain masks or suppresses the distortion so that only an abnormal temperature dependence of the lattice parameters is observed in powders ground from single crystals, with no detectable symmetry change at low temperature. This was true even for gently ground, coarse powders and for powders gently ground under liquid nitrogen. The lattice parameters determined from powders ground from the crystals showed behavior similar to previously reported data for polycrystalline MnBi \cite{Koyama-2008, Yang-2011}.

The grinding-induced strain in powders produced from single crystals could be relieved to some degree by annealing, and diffraction patterns from the annealed powders are shown and discussed in the Supplemental Material. Rietveld analysis of data collected at 20 K from annealed powder indicate the low temperature structure adopts space group $Cmcm$ with lattice constants $a$ = 4.269 {\AA}, $b$ = 7.404 {\AA}, and $c$ = 6.062 {\AA}, in good agreement with those determined from the single crystal measurements shown in Fig. \ref{fig:xray}c, and Mn at (0, $\frac{1}{2}$, 0) and Bi at (0, 0.167(1), $\frac{1}{4}$). This structure closely related to the structure adopted by Mn$_{1.08}$Bi (space group $Pmma$) \cite{Cenzual-1991}, but without the excess Mn in the interstices. As discussed in the Supplemental Material, the $Cmcm$ model is the simplest way to describe the distortion (it is a subgroup of the high temperature space group), and it fits the powder diffraction data well. Thus, it is likely to be the correct low temperature structure. To examine the structure of the low temperature phase further, single crystal x-ray diffraction measurements were performed at 80 and 200 K. The distortion is very small, near the detection limit for laboratory single crystal diffraction measurements, but can be identified with some certainty. Details can be found in the Supplemental Material. The results were consistent with the powder diffraction results. Orthorhombic space groups $Cmcm$, $Cmc2_1$, and $C2cm$ were identified as candidates based on systematic absences. A good structure refinement was achieved in the $Cmcm$, the highest symmetry group of the three candidates and a subgroup of the high temperature space group. The refinement gives Mn at (0, $\frac{1}{2}$, 0) and Bi at (0, 0.16611(6), $\frac{1}{4}$).

\subsection{Neutron diffraction}

\begin{figure}
\begin{center}
\includegraphics[width=3.5in]{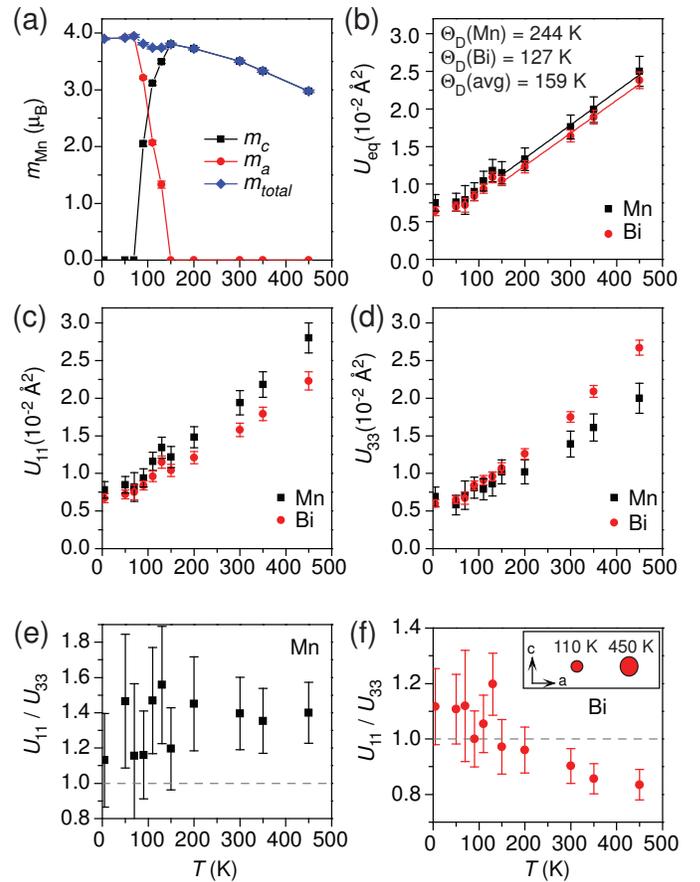}
\caption{\label{fig:neutron}
Results of single-crystal neutron diffraction measurements on MnBi analyzed using the hexagonal NiAs structure type at all temperatures.
(a) Refined magnetic moment on Mn, including the total moment, as well as the projections of the moment along the hexagonal \textit{a}-axis and \textit{c}-axis.
(b) Equivalent isotropic atomic displacement parameters, $U_{eq}$. The element specific and average Debye temperatures and the linear fits used to determine them are shown on the plot.
(c, d) Anisotropic atomic displacement parameters for both atoms. $U_{11}$ measures displacement in the \textit{ab}-plane, and $U_{33}$ measures displacement along the \textit{c}-axis.
(e, f) The ratio $U_{11}/U_{33}$, a measure of the compression of the displacement ellipsoids along the crystallographic c-axis. The inset in (f) compares the relative sizes and shapes of the Bi ellipsoids at 110 and 450 K.
}
\end{center}
\end{figure}

Results of our single crystal neutron diffraction measurements are summarized in Fig. \ref{fig:neutron}. Previously reported neutron diffraction results are limited to powder diffraction \cite{Roberts-1956, Andersen-1967, Yang-2002}. The single crystal data allow us to not only determine with greater reliability the intrinsic nature of the spin reorientation transition, but also to extract information about the lattice vibrations through analysis of the temperature dependence of the atomic displacement parameters.

The neutron diffraction experiment did not have sufficient resolution to detect the small lattice distortion observed by X-ray diffraction (Fig. \ref{fig:xray}). For this reason, the neutron data were analyzed using the hexagonal NiAs-type structure at all temperatures. Thus, for $T \leq$ 90 K, there is potential for systematic effects on the refinement results. Possible effects on the magnetic moment and structural properties derived from the data are discussed below.

The refinement results for the Mn magnetic moment are shown in Fig. \ref{fig:neutron}a. At all temperatures, separate refinements were performed with the moment along \textit{c}-axis, with the moment in the \textit{ab}-plane, and with the moment having projections along both the \textit{c}-axis and in the \textit{ab}-plane. As expected, the results show moments along the \textit{c}-axis at higher temperatures ($\geq$ 150 K)  and in the \textit{ab}-plane at lower temperatures ($\leq$ 70 K). The model with partially rotated moments provided a better fit to the data only for the intermediate temperatures. The in-plane component of the moment is along the \textit{a}-axis.

The magnetic structure with the in-plane components has a lower symmetry (orthorhombic) than that used to describe the nuclear structure (hexagonal). Therefore, three equivalent magnetic domains, generated by rotations of 120 degrees about the $c$-axis, are expected in the large single crystal used for this study. These domains, equally weighted, were used in the refinements of the neutron data. Below the structure transition at 90 K, each magnetic domain can be uniquely related to an orthorhombic structure domain, and the moment along the [100]/[010]/[110] directions in the hexagonal description aligns along the \textit{a}-axis of the orthorhombic lattice of each structural domain (see Fig. \ref{fig:xray}d). Because of this domain formation, we expect little effect on the magnetic structure refinement to arise from use of hexagonal symmetry at low temperature.

Previous studies have reached conflicting conclusions regarding the completeness of the spin rotation at low temperatures, with some finding partial rotation \cite{Roberts-1956, Yang-2002}, and other full rotation \cite{Andersen-1967, Hihara-1970}. Our results show that in MnBi single crystals, the rotation is complete. The observation of only partial rotation in powder samples may be related to the sensitivity of this material to strain. As noted above, the structural phase transition observed at $T_{SR}$ is suppressed with only a small amount of  mechanical stress (gentle grinding). Thus, a strain effect on the spin reorientation itself may be expected.

We find an ordered moment on Mn of 3.50(2) $\mu_B$ at 300 K, and 3.90(2) $\mu_B$ at 5 K. These compare well with the values of 3.6 and 4.1 $\mu_B$ determined from our magnetization measurements at 300 and 5 K, respectively (see below). They are in reasonable agreement with powder diffraction reports as well \cite{Yang-2002}. Note that assignment of nominal oxidation states would give Mn$^{3+}$Bi$^{3-}$, and trivalent Mn (3$d^4$) would be expected to have a moment of $gS$ = 4 $\mu_B$.

Atomic displacement parameters (ADP) determined from the neutron diffraction data are also shown in Fig. \ref{fig:neutron}. These are a measure of the mean squared displacement of atoms from their ideal crystallographic position, and include contributions from thermal motion and, when present, static displacements (off-centering). The data provides not only information about the temperature dependence of the ADPs, but also their anisotropy. Fig. \ref{fig:neutron}b shows the equivalent isotropic displacement parameters $U_{eq}$, while Figs. \ref{fig:neutron}c,d show the anisotropic values. The displacement in the \textit{ab}-plane is measured by $U_{11}$. The displacement along the \textit{c}-axis is measured by $U_{33}$. The ratio $U_{11}/U_{33}$ for each atom is shown in Figs. \ref{fig:neutron}e,f. This ratio characterizes the degree of oblateness of the ellipsoids (squashing along the \textit{c}-axis).

Inspection of $U_{eq}$ (Fig. \ref{fig:neutron}b) shows that both atoms, despite their disparate masses, vibrate similarly in the crystal lattice. This indicates potential wells with similar force constants are experienced by both Mn and Bi. From the temperature dependence, element specific Debye temperatures ($\Theta_D$) can be extracted \cite{Sales-2001}. Above the Debye temperature, and in the Debye model, the slope ($S$) of $U_{eq}$ vs $T$ is given by $S = 3h^2/(4\pi^2mk_B\Theta_D)$, where m is the atoms mass, h is Planck's constant and $k_B$ is Boltzmann's constant. For mass in atomic mass units, and the slope measured in {\AA}$^2$/K, $\Theta_D$ in K is given by $\Theta_D = \sqrt{\frac{146}{m~S}}$. The Debye temperatures determined in this way, and the linear fits used to determine them, are shown on Fig. \ref{fig:neutron}b). The value determined using average mass and average slope is also shown.

Among the results for the anisotropic ADPs, the temperature dependence of the oblateness of the Bi ellipsoids are particularly noteworthy (Fig. \ref{fig:neutron}d). Above about 100 K, $U_{11}/U_{33}$ for Bi decreases steadily up to the highest measurement temperature. That is, the Bi ellipsoid becomes more and more elongated along the \textit{c}-direction as temperature increases, while no temperature dependence in the shape of the Mn ellipsoid can be resolved (Fig. \ref{fig:neutron}e). The Bi ellipsoids are shown in the inset of Fig. \ref{fig:neutron}f for $T$ = 110 K, just above the structural distortion, and 450 K, the highest temperature investigated.

The anisotropy of the Bi ADPs reflects the anisotropy of the restoring force experienced by Bi when it is moved away from its ideal position. This restoring force can arise from both chemical bonding and magnetoelastic effects. While the detailed origin of the ADP anisotropy would require a careful theoretical investigation, it seems reasonable to conclude that the increase in the Bi vibrations along the \emph{c}-axis as temperature is raised contributes to the anisotropic thermal expansion in this material (Fig. 2c). This anisotropic thermal expansion has been linked in a recent theoretical study to the magnetic anisotropy and spin reorientation in MnBi \cite{Zarkevich-2014}. The important role played by the nominally non-magnetic Bi atoms in the magnetism of MnBi also has been recently identified by first principles calculations \cite{Antropov-2014}, where the fine details of the Bi-Bi exchange interactions and spin-orbit coupling on Bi are demonstrated to influence strongly the magnetic anisotropy. Thus, atomic displacement parameters, as especially those of Bi, may provide a fundamental link between the temperature dependence of structural and magnetic properties, and could be a fruitful target for further theoretical investigations.

While $U_{33}$ (Fig. \ref{fig:neutron}d) for both Mn and Bi varies smoothly throughout the entire temperature range, a clear kink in $U_{11}$ (Fig. \ref{fig:neutron}c) is seen for both atoms between 130 and 150 K. This is also reflected in $U_{eq}$ values of Fig. \ref{fig:neutron}b. It is expected that the orthorhombic distortion would manifest itself here as an inflation of the in-plane displacements determined using the hexagonal structure. However, this should appear for $T \leq $ 90 K. The kink in $U_{11}$ near 140 K occurs in the hexagonal structure, and is robust against changes in the magnetic structure model. This is the temperature at which the hexagonal lattice constants experience a deviation from their high temperature behavior (Fig. \ref{fig:xray}c), and the moments begin to rotate away from the \textit{c}-axis (Fig. \ref{fig:neutron}a). This shows that not only the crystal structure and symmetry, but also the atomic displacements in MnBi are closely coupled to the magnetism.

\subsection{Physical properties}

\begin{figure}
\begin{center}
\includegraphics[width=3.25in]{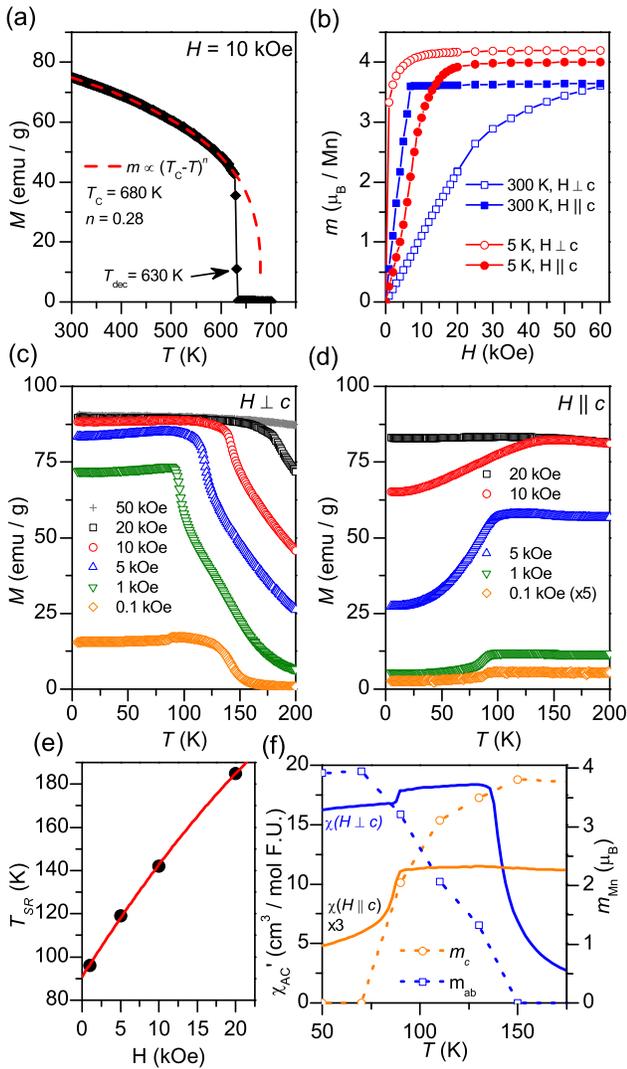}
\caption{\label{fig:mag}
Magnetic properties of MnBi single crystals. (a) Magnetization (\textit{M}) above room temperature. A simple model, shown on the plot, is used to estimate the hypothetical Curie temperature. (b) Magnetic moment (\textit{m}) vs applied field at 5 and 300 K. (c and d) Magnetization vs temperature measured at the indicated magnetic fields for H perpendicular and parallel to the c-axis, respectively. (e) Field dependence of the spin reorientation temperature defined by the minimum of d\textit{M}/d\textit{T} for $H \bot c$. The line is a fit using a second order polynomial. (f) ac magnetic susceptibility measured at 20 Hz in zero applied dc field, with magnetic moment components determined by neutron diffraction shown for comparison.
}
\end{center}
\end{figure}

\begin{figure}
\begin{center}
\includegraphics[width=2.5in]{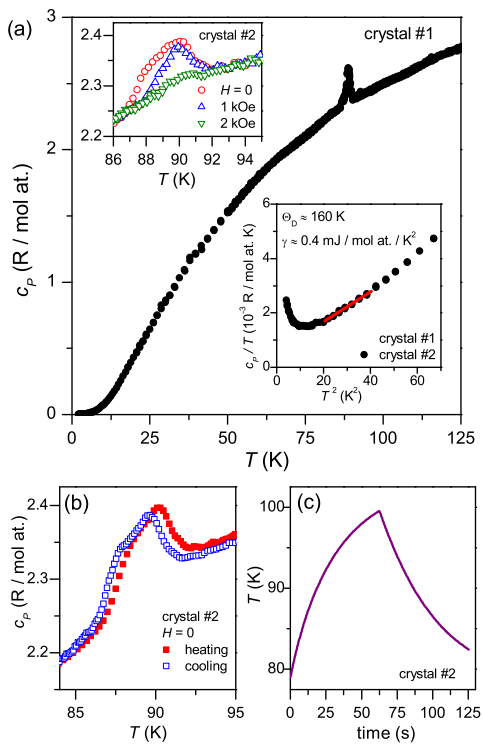}
\caption{\label{fig:hc}
Heat capacity of MnBi single crystals showing an anomaly at the spin-reorientation and crystallographic phase transition at 90 K. The upper inset shows the effects of a magnetic field near $T_{SR}$ and the lower inset shows the low temperature behavior of $c_P / T$ vs $T^2$. Linear fits to the data just above the low-temperature upturn are used to estimate the Debye temperature $\Theta_D$ and the Sommerfeld coefficient $\gamma$. (b) Heat capacity derived from temperature vs. time data collected during a large heat pulse which moves the sample through the phase transition temperature on heating and subsequent cooling, showing the intrinsic thermal hysteresis associated with the transition is $<$ 0.6 K. (c) Temperature vs time data used to determine the $c_P$ data shown in (b).
}
\end{center}
\end{figure}

\begin{figure}
\begin{center}
\includegraphics[width=2.5in]{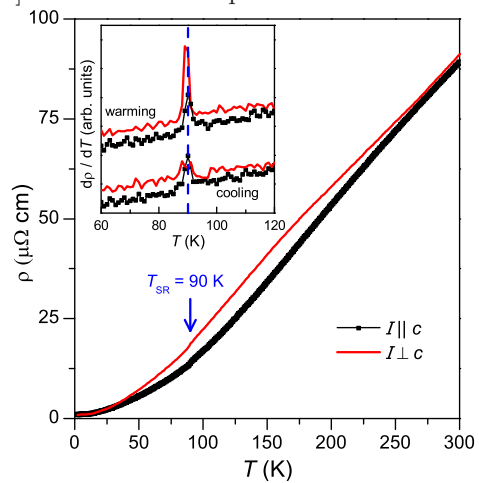}
\caption{\label{fig:res}
Resistivity of MnBi single crystals. Data are shown with current (\textit{I}) flowing both parallel and perpendicular to the c-axis. The upper inset shows the temperature derivative of the resistivity near the $T_{SR}$ measured both upon warming and cooling. The derivative curves are offset vertically for clarity.
}
\end{center}
\end{figure}

Effects of the crystallographic and magnetic phase transitions occurring at $T_{SR}$ on the physical properties of the MnBi single crystals were examined. The magnetic behavior, shown in Fig. \ref{fig:mag}, is consistent with previous literature reports on polycrystalline materials. Detailed reports of the heat capacity and anisotropic electrical resistivity are not available for comparison. Measurements on the single crystals reveal sharp anomalies in both heat capacity (Fig. \ref{fig:hc}) and resistivity (Fig. \ref{fig:res}). The following will discuss the individual properties in some detail.

Results of magnetization measurements on the crystals are summarized in Fig. \ref{fig:mag}. The results are in general agreement with the previously reported behaviors noted in the Introduction. Upon heating above room temperature (Fig. \ref{fig:mag}a), the magnetization decreases and then abruptly drops by orders of magnitude near $T_{dec}$ = 630 K, signaling the decomposition of MnBi. A simple model \cite{Smart} is used to fit the temperature dependence up to $T_{dec}$, giving an estimate of the hypothetical Curie temperature of 680 K.

Isothermal magnetization curves measured with the field parallel and perpendicular to the \textit{c}-axis are shown in Fig. \ref{fig:mag}b. This data demonstrates the change in the easy axis of magnetization from along \textit{c} near room temperature to perpendicular to \textit{c} at 5 K. The spin reorientation is also clearly seen in the temperature dependent magnetization results shown in Fig. \ref{fig:mag}c,d. Upon cooling in the lowest applied fields, the magnetization for $H \bot c$ increases below about 150 K, and the magnetization for $H \| c$ decreases abruptly below 90 K. This is consistent with previous reports from textured polycrystalline material \cite{Liu-2005} and with our neutron diffraction data discussed above, which find the moments to rotate away from \textit{c} below about 140 K and lie fully in the \textit{ab}-plane below 90 K. The dc magnetic susceptibility (\textit{M/H} at low field) is low when measured perpendicular to the moment direction when the moments are locked into the \textit{c} direction or the \textit{ab}-plane. At intermediate temperatures, when the moment has components both parallel and perpendicular to \textit{c}, the magnetic susceptibility is large in both directions. This is clearly illustrated in in Fig. \ref{fig:mag}f, which compares ac magnetic susceptibility measured at $H_{dc}$ = 0 with the magnetic moment components determined by neutron diffraction.

For $H \bot c$, a small but sharp decrease in $m$ is observed at 90 K for H = 0.1 kOe (also seen in the zero field ac measurement in Fig. \ref{fig:mag}f). Such a feature could occur from misalignment of the crystal during the measurement; however, its absence in the 1 kOe data suggests it is an intrinsic and field dependent feature. At \textit{H} = 1 kOe in this direction, a sharp increase in $m$ is observed near 100 K. Signatures of the spin reorientation transition in the magnetization data are generally broadened and moved to higher temperatures as the applied magnetic field is increased. Taking the point of minimum slope in the $m(T)$ data for $H \bot c$ and 1 kOe $\leq H \leq$ 2 kOe as $T_{SR}$, the temperature at which the spin reorientation is complete, a plot of $T_{SR}(H)$ is shown in Fig. \ref{fig:mag}e. Interpolating this back to $H$ = 0 using a second order polynomial gives $T_{SR}$(0) = 90 K, coincident with sharp drops in $\chi_{ac}$ (Fig. \ref{fig:mag}f) and $m$ measured at $H$ = 0.1 kOe (Fig. \ref{fig:mag}c). A simple linear fit to the data in Fig. \ref{fig:mag}e gives a similar value of $T_{SR}$(0) = 95 K.

The measured heat capacity $c_P$ of MnBi single crystals is shown in Fig. \ref{fig:hc}. Measurements are shown for two different crystals. A sharp thermal anomaly is observed at $T_{SR}$ = 90 K. A broad feature has been reported in the heat capacity of polycrystalline MnBi at a significantly higher temperature of 118 K \cite{Kondo-1950}. The integrated entropy associated with the peak for $H$ = 0 is 0.01 R (mol Mn)$^{-1}$. Similar anomalies are seen in the heat capacity of Nd$_2$Fe$_{14}$B and related compounds have been reported, and used as a precise way to determine spin reorientation temperatures in these materials \cite{Fujii-1987, Pique-1996}. The heat capacity anomaly at the spin reorientation in MnBi is suppressed by applied magnetic fields, as shown in the upper inset of Fig. \ref{fig:hc}a, decreasing in magnitude with increasing field while remaining at 90 K.

There is a low temperature upturn in $c_P$ upon cooling below about 2.5 K, shown in the lower inset of Fig. \ref{fig:hc}(a). The origin of this behavior is unclear at this time. No previous reports of the low temperature heat capacity for MnBi were found in the literature for comparison. To extract a Debye temperature for comparison with those determined from the temperature dependence of the ADP values discussed above (Fig. \ref{fig:neutron}e), linear fits to the low-temperature $c_P/T$ vs $T^2$ data were performed, excluding the upturn below 2.5 K. The values of the Debye temperature and Sommerfeld coefficients determined from them are listed in the Figure. The Debye temperature determined in this way is in very good agreement with the average Debye temperature of 159 K determined from the neutron diffraction data.

To examine the behavior of the heat capacity near $T_{SR}$ in detail, measurement were performed in which a large heat pulse is applied to the sample and then removed, so that the sample passes through the spin reorientation upon heating and then again upon cooling. Analysis of the time dependence of the sample temperature \cite{Lashley-2003} is used to derive heat capacity values, using software provided by Quantum Design. Fig. \ref{fig:hc}b shows the heat capacity determined separately from the heating and cooling curves. A thermal hysteresis in the peak position of about 0.6 K is observed. Since there is expected to be some thermal lag between the sample and the thermometer in this measurement, this can be considered an upper bound any the intrinsic thermal hysteresis. Inspection of the temperature vs. time curves (Fig. \ref{fig:hc}c) used to determine the heat capacity reveals no detectable thermal arrest near $T_{SR}$, indicating little latent heat is associated with the transition.

Fig. \ref{fig:res} shows results of electrical resistivity measurements on MnBi single crystals. For the current parallel and perpendicular to the \emph{c}-axis, the residual resistivity ratios [$\rho$(300 K) / $\rho$(5 K)] are 88 and 98, respectively. A decrease in resistivity is observed upon cooling through $T_{SR}$, indicated by the arrow on the figure. This is shown clearly by the temperature derivative shown in the upper inset of Fig. \ref{fig:res}. Hihara and Koi observed a small resistivity anomaly in polycrystalline MnBi occurring near 115 K \cite{Hihara-1970}. Resistivity measurements on thin films have shown anomalies near 50 K \cite{Kharel-2011}. The anomaly is most clearly seen in d$\rho$/d$T$ shown in the inset of Fig. \ref{fig:res}. No thermal hysteresis is observed within the 1 K resolution of the resistivity data. Due to the irregular shape of the crystals, there is some uncertainty associated with the absolute value of the resistivity reported here. However, the data does suggest that there is not a large amount of anisotropy in the electrical transport. The material is a good metal, reaching about 1 $\mu\Omega cm$ at 2 K. Measurements in the \emph{ab}-plane performed on films have shown similar resistivity values at near room temperature, but larger low temperature resistance than we find in our crystals \cite{Kharel-2011}.

\section{Summary and conclusions}

In summary, we have identified a symmetry-lowering structural phase transition which occurs at the spin-reorientation temperature in MnBi, driven by magnetostriction. Below $T_{SR}$ = 90 K, when the magnetic moment lies in the \textit{ab}-plane the structure is orthorhombic. Diffraction data indicates that the low temperature structure can be described in space group $Cmcm$, a subgroup of \textit{P}6$_3$/\textit{mmc}, the space group of the hexagonal NiAs-type structure adopted at higher temperature when the moment in aligned along the \textit{c}-axis. Such a distortion is allowed to occur via a second order transition. Heat capacity data place an upper bound of 0.6 K on the thermal hysteresis of the spin reorientation. Magnetoelastic coupling results in a shortening of the axis along which the Mn moments are directed in both phases. Single crystal neutron diffraction analysis reveals an increasing elongation of the Bi displacement ellipsoid along the \textit{c}-axis as temperature is increased in the high temperature state. This is likely related to the observed anisotropic thermal expansion, and may play an important role in the magnetocrystalline anisotropy, which has been attributed to Bi$-$Bi exchange interactions \cite{Antropov-2014}. The relationship between atomic vibrations and magnetic anisotropy is also supported by the kink observed in the atomic displacement parameters at 140 K, a temperature well above the orthorhombic distortion, and at which the moment first begins to cant away from the \textit{c}-axis upon cooling. The crystallographic phase transition reported here may prove to be key in interpreting future experimental and theoretical studies of the intrinsic properties of this unusual permanent magnet material.
\\

Research sponsored by the U. S. Department of Energy, Office of Energy Efficiency and Renewable Energy, Vehicle Technologies Office, Propulsion Materials Program (M.A.M.) and the Critical Materials Institute, an Energy Innovation Hub funded by the U.S. Department of Energy, Office of Energy Efficiency and Renewable Energy, Advanced Manufacturing Office (B.C.S.).  Neutron diffraction measurements conducted at ORNL's High Flux Isotope Reactor (H.C. and B.C.C.) were sponsored by the Scientific User Facilities Division, Office of Basic Energy Sciences, US Department of Energy. The authors thank David S. Parker and David J. Singh for helpful discussions and insight, and for suggesting that a structural distortion should occur in this material, and Bayrammurad Saparov and Radu Custelcean for assistance with transport measurements and single crystal x-ray diffraction, respectively.


\end{document}